\documentclass[aps,prb,twocolumn,showpacs,superscriptaddress]{revtex4-1}
\usepackage{graphicx}

\begin{document}

\title{Neutron scattering study of magnetic phase separation in nanocrystalline La$_{5/8}$Ca$_{3/8}$MnO$_3$}

\author{Chetan Dhital}
\affiliation{
Department of Physics, Boston College, Chestnut Hill, Massachussetts 02467, USA
}

\author{Clarina de la Cruz}
\affiliation{Neutron Scattering Science Division, Oak Ridge National Laboratory, Oak Ridge, Tennessee 37831-6393, USA}

\author{C. Opeil}
\affiliation{
Department of Physics, Boston College, Chestnut Hill, Massachussetts 02467, USA
}
\author{A. Treat}
\affiliation{
Department of Physics, Boston College, Chestnut Hill, Massachussetts 02467, USA
}
\author{K.F. Wang}
\affiliation{
Department of Physics, Boston College, Chestnut Hill, Massachussetts 02467, USA
}
\affiliation{Nanjing National Laboratory of Microstructures and Department of Physics, Nanjing University, Nanjing 210093, China} 

\author{J.-M. Liu}
\affiliation{Nanjing National Laboratory of Microstructures and Department of Physics, Nanjing University, Nanjing 210093, China} 

\affiliation{International Center for Materials Physics, Chinese Academy of Sciences, Shenyang 110016, China}

\author{Z.F. Ren}
\affiliation{
Department of Physics, Boston College, Chestnut Hill, Massachussetts 02467, USA
}
\author{Stephen D. Wilson}
\email{stephen.wilson@bc.edu}
\affiliation{
Department of Physics, Boston College, Chestnut Hill, Massachussetts 02467, USA
}

\begin{abstract}
We demonstrate that magnetic phase separation and competing spin order in the colossal magnetoresistive (CMR) manganites can be directly explored via tuning strain in bulk samples of nanocrystalline La$_{1-x}$Ca$_x$MnO$_3$.  Our results show that strain can be reversibly frozen into the lattice in order to stabilize coexisting antiferromagnetic domains within the nominally ferromagnetic metallic state of La$_{5/8}$Ca$_{3/8}$MnO$_3$.  The measurement of tunable phase separation via magnetic neutron powder diffraction presents a direct route of exploring the correlated spin properties of phase separated charge/magnetic order in highly strained CMR materials and opens a potential avenue for realizing intergrain spin tunnel junction networks with enhanced CMR behavior in a chemically homogeneous material.        
\end{abstract}


\maketitle

\section{Introduction}
Phase separation is widely believed to play a prominent role governing the fundamental properties of the colossal magnetoresistive (CMR) manganites \cite{dagottoreview, ahn, burgy, MoreoScience} where a coexistence of energetically similar electronic states can be stabilized via perturbations of the lattice, spin, or charge ordering. This has been demonstrated vividly in highly strained thin CMR films where lattice strain fields stabilize phase separated domains possessing dramatically different electronic properties \cite{biswas, fath}. A fundamental limitation however in understanding the detailed magnetic properties within these strain-induced, phase separated, domains stems from an inability to effectively tune strain in bulk samples where spin sensitive, momentum-resolved probes such as neutron diffraction can be leveraged.  Recent work however exploring the tunable properties of nanocrystalline CMR materials, in particular La$_{1-x}$Ca$_x$MnO$_3$ (LCMO) \cite{sarkar,kharlamova,rozenburg,muroi,mahesh,rivas,li,hueso}, suggests an alternate path to realizing high strain fields in large volumes of CMR material and thereby potentially rendering phase separated magnetic order accessible to neutron scattering studies.    

Dramatic changes in the properties of spin systems are known to appear in materials reduced to the nanoscale \cite{batlle} where both finite size, grain boundary, and surface strain effects can play an important role. Among these influences, strain effects seem to play a dominant role in nanocrystalline LCMO where the effect of surface strain in nanocrystalline grains substantially influences magnetic phase formation. In particular, the ordered moment within the ferromagnetic regime is suppressed due to the growth of a magnetic dead-layer along the strained outer shell of crystallite boundaries \cite{curial,bibes,quintela,muroi}.   While the detailed interactions within this surface layer remain debated, one possible origin is an enhanced competition between ferromagnetic (FM) and antiferromagnetic (AF) exchange interactions along the strained surface region. This suggests a promising avenue of utilizing strain in order to explore both the core role of phase separation in CMR materials as well as a further route for tunable material applications. One of the key unanswered questions however remains whether strain in nanocrystallites can be tuned and enhanced further in order to completely stabilize competing electronic phases in a manner similar to that employed in highly strained thin films \cite{biswas, fath}.

In this paper, we present neutron diffraction measurements demonstrating the first observation of strain-induced magnetic phase separation within LCMO nanocrystallites.  By mechanically stressing nominally ferromagnetic La$_{5/8}$Ca$_{3/8}$MnO$_3$ nanocrystallites further through high-energy ball milling techiques, we observe the appearance of an anisotropically enhanced strain field coupled to the emergence phase separated AF order. The resulting out-of-plane b-axis of LCMO nanograins is compressed leading to an enhanced Mn-O-Mn superexchange pathway, and as a result, an antiferromagnetic state is stabilized that coexists with the competing ferromagnetic order present in bulk, strain-free samples.  Our results provide a new avenue for probing the nature of phase separation and phase competition in the LCMO system and suggest a potential route toward realizing a spin tunnel junction network with an enhanced magnetoresistive response \cite{yunoki} within a chemically homogenous material.    

\section{Experimental Details}
For our experiments, we synthesized nanocrystalline La$_{5/8}$Ca$_{3/8}$MnO$_{3}$ powder by mixing stoichiometric amounts of La$_{2}$O$_{3}$ (4N), CaCO$_{3}$ (3N), and MnCO$_3$ (3N) and firing them in air at 1400 $^\circ$C with five intermediate grinding and mixing steps at this same temperature.  The final resulting powder was then pressed and sintered in air at 1400 $^\circ$C for 24 hours and then ground inside a SPEX 8000 high energy ball mill for 24 hours.  A separate annealed sample was also created from a portion of this nanopowder by then hot pressing it at 1000 $^\circ$C under 100 MPa of pressure for 3 minutes.  Neutron experiments were performed on the HB-2A powder diffractometer at the High Flux Isotope Reactor at Oak Ridge National Lab.  A $\lambda_i$=$1.5385 \AA$ was used with a Ge(115) monochromator and $12'$-$31'$-$6'$ collimation, and data were refined using the FullProf software package.  Crystallite imaging was performed on a JEOL 6340F scanning electron microscope (SEM), and magnetization data was collected in an Oxford MagLab measurement system.  Resistivity data was collected within a Quantum Design PPMS system.  

\section{Neutron Scattering and Magnetoresistance Results}
Initially, the resistance of the starting, bulk polycrystalline La$_{5/8}$Ca$_{3/8}$MnO$_3$ powder was measured in both 0 T and 7 T from room temperature down to $4$ K.  The resulting raw resistivity data and the magnetoresistance (MR) ratio are plotted in Figs. 1 (a) and 2 (d) respectively.  Resistivity as well as the MR ratio data from this bulk polycrystalline sample agree well with previous reports and peak near the expected ferromagnetic $T_c$---this serves as a useful starting reference for analyzing transport results obtained from subsequent nanocrystalline samples engineered from this same batch of powder.  A portion of this bulk starting powder was then ball-milled into the nanocrystalline phase as described in Section II, and the resulting grain sizes were analyzed via SEM measurements.  Looking at SEM images in Fig. 2 (a) for the as-milled La$_{5/8}$Ca$_{3/8}$MnO$_3$ nanopowder, scans reveal average grain sizes ranging from $\approx 100-200$ nm and confirm a substantial reduction from the initial bulk polycrystalline state.  EDS measurements were also taken across a number of different regions on this nanocrystalline specimen and show the final Ca concentration of this sample to be x$=0.38\pm0.02$, verifying a chemically homogeneous sample.     

The zero field resistivity of this nanopowder sample was measured after cold pressing a pellet under $60,000$ psi, and the magnetoresistance was measured under a 9T field following zero field cooling (ZFC). The resulting resistance data is plotted in Fig. 1 (b) and the directly subtracted R(0T)-R(9T) data are plotted in Fig. 2 (c).  If one assumes that the physical connectivity between grains is field independent, this subtraction removes the extrinsic effect of variable distance between nanocrystallite grains upon cooling and reveals two distinct peaks---the expected magnetoresistive (MR) peak at T$_C$ and a second anomalous peak at 140 K. Secondary broad MR peaks have been reported previously as a reflection of spin polarized intergrain tunneling at disordered interfaces in bulk LCMO powders \cite{hong,yuan}; however our subsequent neutron diffraction results instead suggest the influence of coexisting short-range AF fluctuations where the maximal MR difference occurs precisely at T$_N$.  The fractional MR change is also over plotted with this difference in Fig. 2 (c) for reference.  It is worth noting that at temperatures just below room temperature, small differences between 0T and 9T resistance data also appear in the MR ratio; however these small fluctuations in the data are likely simply due to noise amplified in taking the MR ratio of data superimposed with a high degree of extrinsic, intergrain, tunneling resistance.  No phase transitions associated with this noise in the MR ratio were identified in our subsequent neutron measurements.

\begin{figure}[t]
\includegraphics[scale=.4]{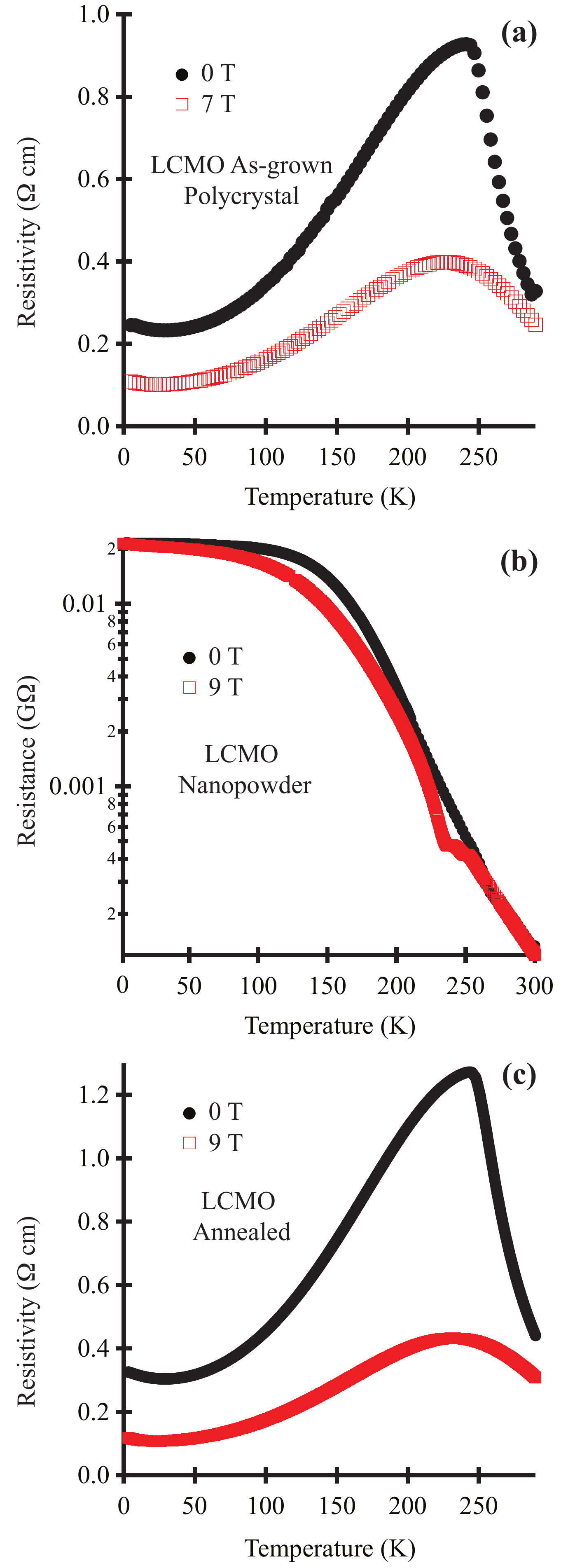}
\caption{Raw resistivity and resistance measurements of (a) the as-grown polycrystalline La$_{5/8}$Ca$_{3/8}$MnO$_3$ powder, (b) the ball-milled La$_{5/8}$Ca$_{3/8}$MnO$_3$ nanopowder from the same batch, and (c) the subsequently hot-pressed annealed La$_{5/8}$Ca$_{3/8}$MnO$_3$ powder in both $0$ T and high magnetic field.  Field measurements were taken following a zero field cooling of the sample.}   
\end{figure}

In order to explore both the structural and magnetic properties of the nanocrystalline La$_{5/8}$Ca$_{3/8}$MnO$_{3}$ sample, we measured the neutron powder diffraction profiles at a variety of temperatures.  Fig. 3 shows both the high temperature (300 K) and low temperature (10 K) diffraction patterns where refinement of the 300 K data within the paramagnetic state was accomplished by utilizing a single nuclear $Pnma$ crystallographic phase with spherical size parameters in order to account for the broadened diffraction widths. Looking first at the lattice constants at 300 K plotted in Fig. 4 (a), refinement reveals a compressed out-of-plane b-axis with b(300K)$=7.6665\pm0.0008 \AA$ that remains unchanged upon cooling to 10 K with b(10K)$=7.6640\pm.0008 \AA$.  Lattice constants evolve continuously with reducing temperature (Fig. 4 (a)), and the unit cell volume remains roughly frozen with V(300K)$=230.36\pm0.06 \AA^3$ and V(10K)$=229.21\pm0.06 \AA^3$ similar to the strain arrested unit cell in nanocrystalline La$_{0.5}$Ca$_{0.5}$MnO$_3$ \cite{chatterji}.      

Both the in-plane OS$_{\parallel}=\frac{2(c-a)}{(c+a)}$ and out-of-plane OS$_{\perp}=\frac{2(a+c-b\sqrt{2})}{(a+c+b\sqrt{2})}$ orthorhombic strain parameters for this sample are plotted in Fig. 4 (b).  Comparing these values to those previously reported in both nanocrystalline and bulk LCMO powders \cite{sarkar}, it is immediately evident that the OS$_{\perp}$ field in this nanocrystalline sample is $\approx4$ times higher than expected.  The in-plane strain field however agrees with previous studies of unstrained LCMO suggesting that the ball milling process anisotropically enhances the OS$_{\perp}$ strain field---consistent with the expected anisotropic compressibility of LCMO \cite{kozlenko}.  As the sample is cooled from 300 K, OS$_{\perp}$ slowly relaxes until T$\approx140$ K is reached and upon further cooling OS$_{\perp}$ remains roughly constant.  There is negligible variation in OS$_{\parallel}$ across all temperatures similar to previous studies of nanophase LCMO \cite{chatterji}.  

\begin{figure}[t]
\includegraphics[scale=.3]{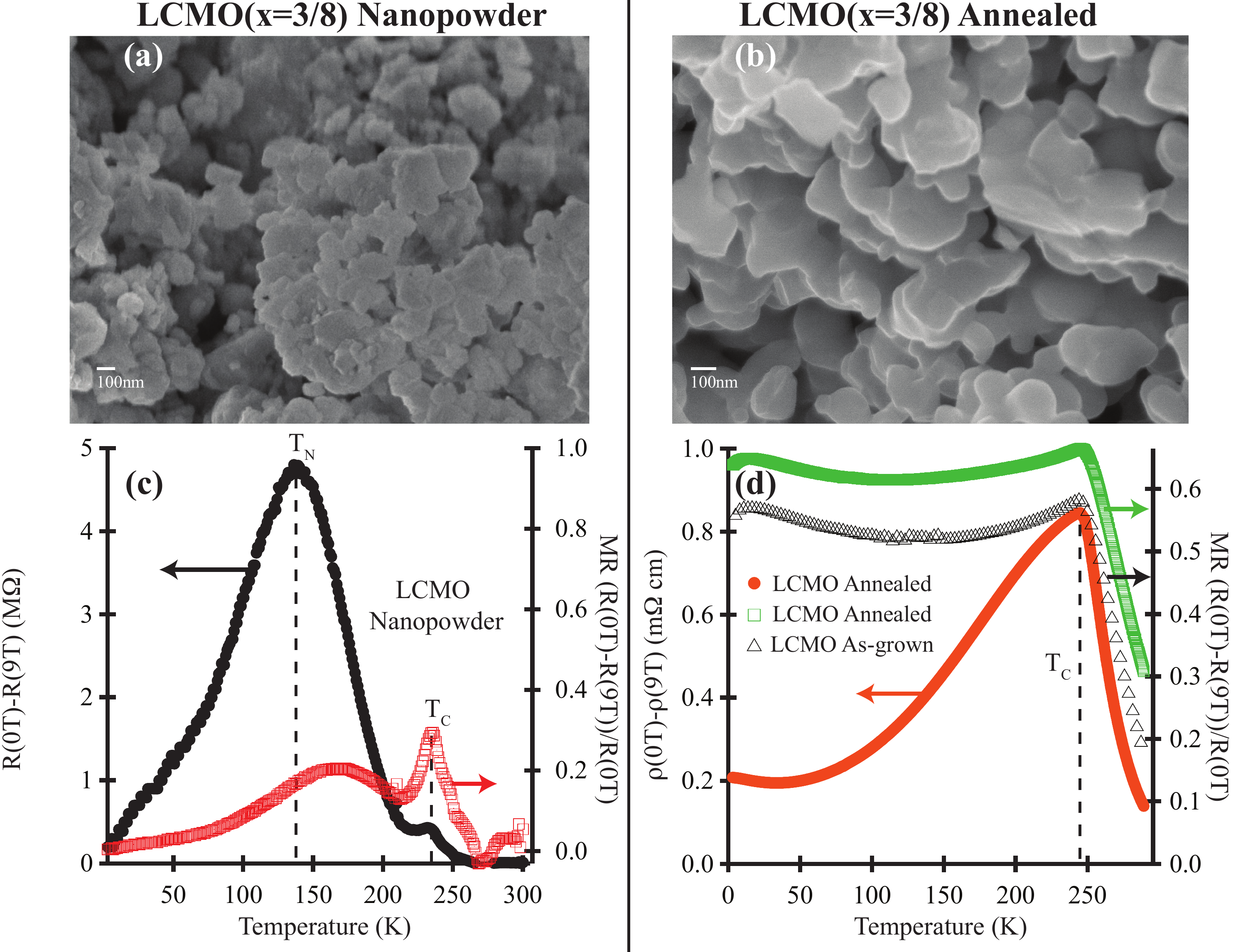}
\caption{(a) SEM image of La$_{5/8}$Ca$_{3/8}$MnO$_3$ nanocrystalline powder sample. (b)  SEM image of annealed La$_{5/8}$Ca$_{3/8}$MnO$_3$ nanopowder sample following hot pressing at 1000 $^\circ$C. Field subtracted resistance and MR plots at 0 T and ZFC 9 T fields are plotted for the (c) as-milled nanopowder LCMO sample and for (d) starting sintered powder and annealed samples respectively. Dashed lines denote T$_c$ and T$_N$ magnetic phase transitions in each sample.}   
\end{figure}

\begin{figure}[t]
\includegraphics[scale=.3]{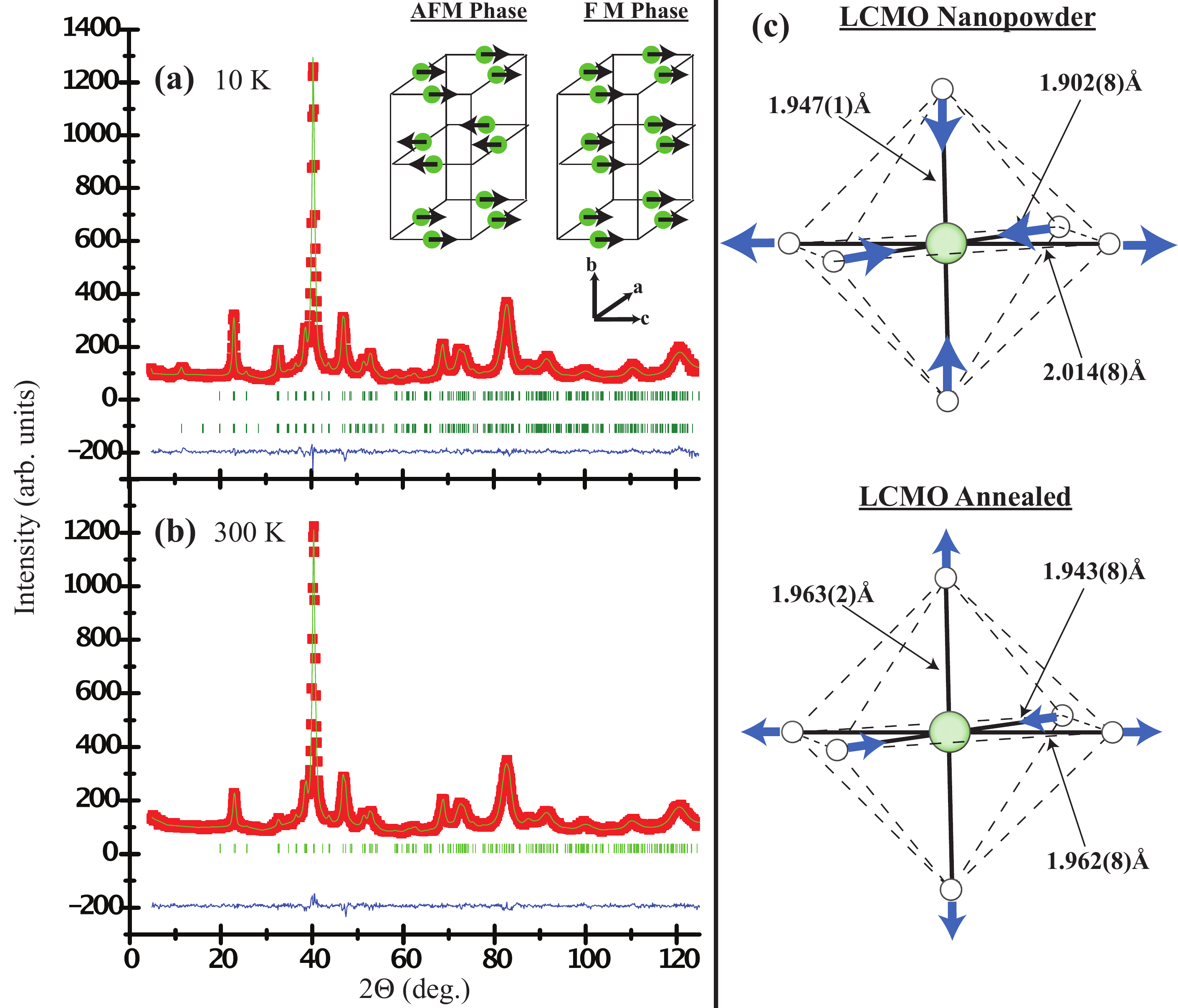}
\caption{Neutron powder diffraction pattern for the La$_{5/8}$Ca$_{3/8}$MnO$_3$ nanopowder sample at (a) 10 K and (b) 300 K.  Inset of (a) shows moment alignments modeled in the AF and FM spin phases.  (c)  Octahedral distortion of both the as-milled nanopowder (top) and annealed samples (bottom).  Out-of-plane Mn-O-Mn bond angles at 250 K are $159.24\pm0.05^{\circ}$ and $158.12\pm0.07^{\circ}$ for the nanopowder and annealed samples respectively.}
\end{figure}

In order to examine the influence of this enhanced strain on the magnetic properties of this system, we turn now to analyzing the spin order probed within our neutron experiments. In agreement with bulk magnetization measurements (Fig. 4(c) inset), FM spin ordering was observed upon cooling below T$_{c}=233$ K, and the FM ordered moment is plotted as a function of temperature in Fig. 4 (c).  The broadened transition is consistent with previous reports suggesting a crossover to a second order magnetic phase transition in LCMO nanocrystallites \cite{sarkar, li}; however the size-induced reduction in the saturated moment to $2.03\pm0.04 \mu_{B}$ is larger than expected given the known variation of ordered moment with particle size \cite{tang, muroi}.  The additional reduction in moment instead points to a fraction of spins participating in a competing, phase separated, region of the nanocrystallites.   

Looking again at Figs. 3(a) and (b), additional reflections appear at low-Q in the 10 K diffraction profile of this nanopowder indicative of the appearance of a coexisting AF phase best modeled as A-type AF order.  Due to the broad nature of these new magnetic reflections and the relatively small ordered moment, the precise moment direction could not be determined; however the moment orientation depicted in the inset of Fig. 3 (a) depicts one possibility using the known A-type spin orientation in the undoped parent system of the LCMO phase diagram.   Fig. 4 (d) shows an expanded view of the lowest order AF peak both above and below T$_N$ at 140 K and 10 K respectively where the peak width at 10 K reveals an AF correlation length of $\zeta$(AF)$=110 \AA$, shorter than that of the FM phase with $\zeta$(FM)$=212 \AA$.   The order parameter for this new AF phase is plotted in Fig. 4 (c) and shows a T$_N\approx140$ K with an ordered moment of $1.06\pm0.03 \mu_{B}$ at 10 K (refer to Section IV of this paper for interpretation of this moment value).   A number of systematic studies examining size effects on magnetism in nanocrystalline LCMO powders have failed to detect this second competing phase \cite{sarkar, tang} which strongly suggests that its stabilization is coupled to the enhanced OS$_{\perp}$ strain field in our ball-milled sample.  The onset of AF order coincides with the arrest in the relaxation of OS$_{\perp}$ plotted in Fig. 4 (b) indicative of a strong coupling between the new AF order and the lattice.

Exploring the influence of this strain field further, we annealed a nanopowder sample taken from the original batch of LCMO nanopowder by quickly hot pressing it in order to anneal strain effects while retaining reduced crystallite size.  The SEM image in Fig. 2 (b) shows that although appreciable grain growth occurred during the annealing process, crystallite sizes remained in the range $\approx300-500$ nm.  Resistivity measurements plotted in Figs. 1 (c) and 2 (d) reveal the disappearance of the secondary MR peak observed within the strained nanopowder with only the expected MR peak remaining at T$_c$. Refining the high temperature powder diffraction profile of this annealed sample revealed a relaxed b-axis lattice parameter of b(250K)$=7.7193\pm0.0007 \AA$ due to the removal of the frozen, milling-induced, strain with the in-plane lattice constants of a(250K)$=5.4532\pm0.0003 \AA$, c(250K)$=5.4855\pm0.0004 \AA$ slightly contracted relative to the earlier nanopowder values of a(260K)$=5.4599\pm0.0007 \AA$, c(260K)$=5.4985\pm0.0008 \AA$, and  b(260K)$=7.6637\pm0.0009 \AA$.  The resulting strain fields (plotted in Fig. 4 (b)) substantially differ from those of the strained as-milled nanopowder.  In particular, the out-of-plane OS$_{\perp}$ field relaxed back to the expected bulk value of OS$_{\perp}\approx0.0025$ \cite{sarkar} while the in-plane OS$_{\parallel}$ reduced only slightly.  For example, at T$=260$K values for OS$_{\parallel}=0.0070\pm0.00019$ and OS$_{\perp}=0.0110\pm0.00015$ in the nanopowder sample relaxed to OS$_{\parallel}=0.0059\pm0.00009$ and OS$_{\perp}=0.0020\pm0.00010$ at T$=250$ K in the annealed sample.  Additionally, the low temperature data from this annealed sample showed only the expected ferromagnetic spin phase with no competing AF order resolvable down to 4 K (see fig. 4 (d)).  The saturated ferromagnetic moment at 4 K increased to $2.78\pm0.04 \mu_B$ revealing a recovery of spins into the ordered ferromagnetic phase at the expense of the destabilized AF order; however a measurable fraction of spins seem to remain disordered.     

\begin{figure}[t]
\includegraphics[scale=.25]{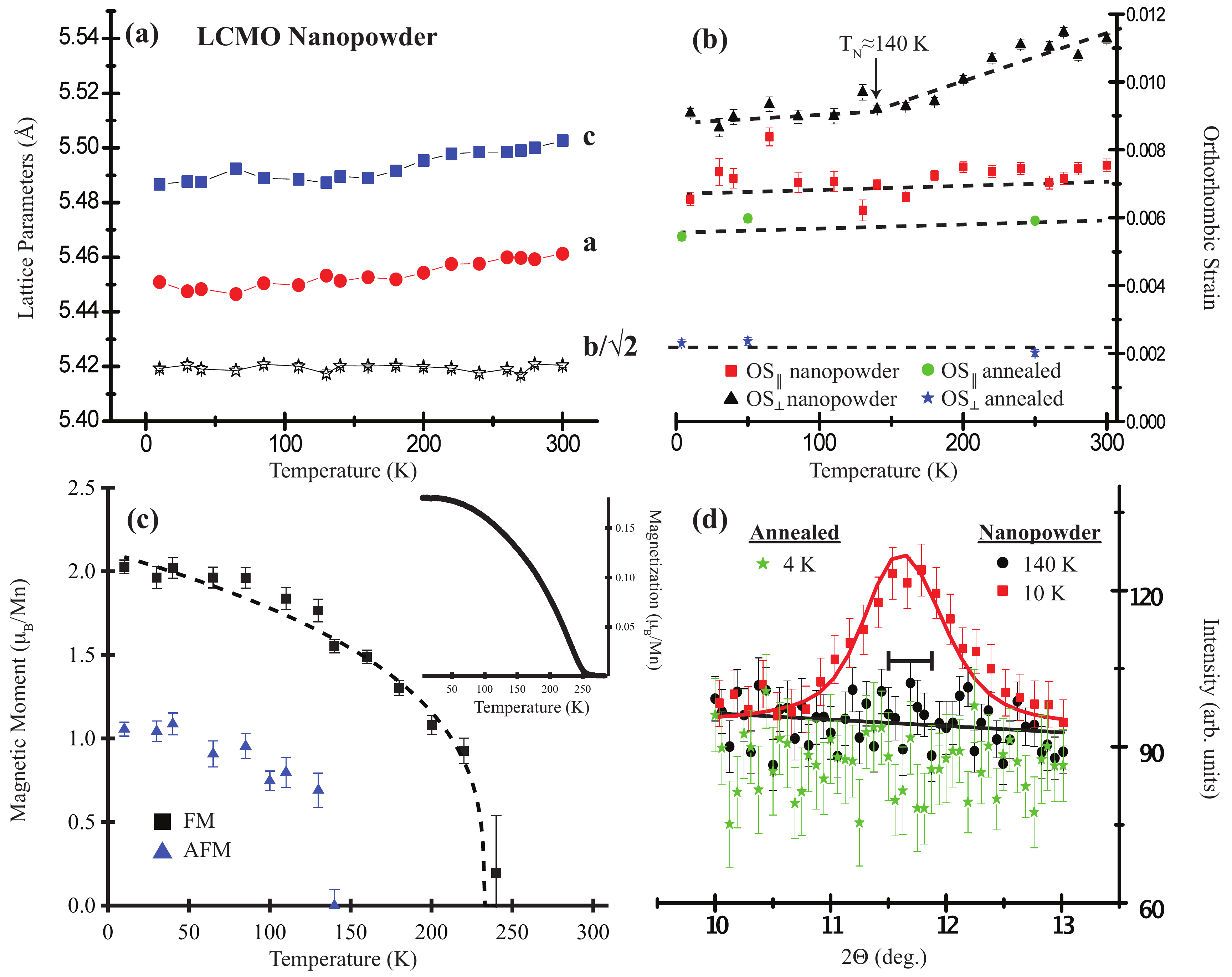}
\caption{(Color online) (a) Lattice parameters for the La$_{5/8}$Ca$_{3/8}$MnO$_3$ nanocrystalline sample. (b)  OS$_{\perp}$ and OS$_{\parallel}$ as a function of temperature in the as-milled nanopowder and annealed samples.  (c) Temperature dependence of the FM and AF order parameters in the as-milled nanopowder sample.  The dashed line is a power law fit to the FM order parameter with $\beta=0.33\pm0.03$.  The inset shows bulk magnetization measurements on the nanopowder in $0.01$ T FC.  (d) The (010) AF reflection below (10 K) and above (140 K) T$_N$ for the as-milled nanopowder and the equivalent scan at 4 K for the annealed powder.  The central bracket denotes the full width at half maximum (FWHM) of the instrument resolution at this peak position. }
\end{figure}  

\section{Discussion}    
Our results demonstrate that anisotropically enhanced strain fields within LCMO nanocrystallites can be utilized to explore phase separation in the manganites and to stabilize competing AF order.  While recent studies of intrinsically phase separated (La$_{1-y}$Pr$_{y}$)$_{1-x}$Ca$_{x}$MnO$_3$ have shown a coupling of microstrain to the relative magnetic volume fractions\cite{pomja}, here our results demonstrate that strain suprisingly stabilizes magnetic phase seperation in a chemically single-phase, nanocrystalline La$_{5/8}$Ca$_{3/8}$MnO$_3$ sample. The b-axis Mn-O-Mn bond lengths at 250 K for the strained nanopowder increase after annealing and slightly increase the bond angle along this exchange pathway (see Fig. 3 (c)) suggesting that the increased OS$_{\perp}$ orthorhombic strain and the resulting shortened b-axis directly enhance the out-of-plane Mn-O-Mn superexchange.  

The Jahn-Teller distortion parameter  $\sigma_{JT}=(\frac{1}{3}\sum((Mn$-$O)$-$<Mn$-$O>)^2)^{1/2}$ at 250 K is dramatically enhanced in the nanopowder sample ($\sigma_{JT}=0.046\pm0.008$) relative to the annealed sample ($\sigma_{JT}=0.009\pm0.004$) (Fig. 2(c)), and $\sigma_{JT}$ remains unchanged upon cooling.  This enhanced Jahn-Teller distortion accompanies the stabilization of phase separated A-type AF order \cite{singh}; however our current measurements do not resolve any direct change in this enhanced $\sigma_{JT}$ in response to the onset of AF order.  This may be due to our neutron measurements' averaging over the entirety of the sample's bond lengths, making subtle changes from within the AF volume fraction difficult to resolve. 

The total ordered moment (AF+FM) in this mixed phase nanopowder is $\approx94\%$ of the known ordered moment in bulk crystalline samples \cite{uehara} in contrast to the annealed sample whose total moment is only $\approx84\%$ of the expected value. Given that the reduced ferromagnetic moment in LCMO nanocrystallites is known to stem from disordered spins at the grain surface \cite{curial, bibes}, this suggests that a substantial portion of spins disordered at grain boundary surfaces in these nanocrystallite grains condense into the AF state under the enhanced strain field observed in our study. The relative volume fractions of the AF and FM phases are difficult to discern in the measurements presented here. Ordered moments from both phases were determined by indexing them to the same scale factor of the nuclear only scattering phase ie. they were assumed to scatter from the entirety of the nuclear scattering volume.  Hence the quoted moments for each phase are meaningful only in terms of a total ordered moment from both phases, but the quoted ordered moments for the individual phases do reflect a minimum estimate for the moments in those sample regions.

If we assume that the ferromagnetic ordered moment of the nanopowder sample should equal that of the annealed sample (when magnetic phase separation disappears), then FM occupies roughly $50\%$ of the nanopowder sample and the remaining $50\%$ consists of the magnetically phase separated AF phase. This picture would then render an AF ordered moment of $\approx1.46 \mu_B$ in the magnetically phase separated regions. This naive picture however ignores the possibility of some fraction of disordered spins at the grain interfaces and should only serve as a rough approximation.            

Our data point toward the creation of a bulk network of LCMO nanocrystallites separated through a series of AF-FM domain walls where spin polarized tunneling plays a dominant role in conduction \cite{yunoki} similar to that envisioned via composites incorporating insulating CMR parent phases \cite{zhu}.  Our present results however only suggest a qualitative realization of this network of AF-FM domains in a chemically homogeneous system, and it currently remains unclear whether relatively coarse ball-milling techniques can be leveraged to tune the anisotropic strain fields in LCMO in a controllable fashion.  While our neutron scattering results only report a bulk average of the samples studied, the onset of short-range AF order in these strained nanocrystallites likely occurs at the grain surfaces where the strain fields are naively the largest and coincides with the appearance of a second MR channel in this system---highlighting the importance of short range AF fluctuations in the MR response of CMR materials with a high density of grain boundaries. This second MR peak in resistivity and the emergence of AF order are not likely to originate from a simple picture of a bimodal size distribution of LCMO grains within our nanopowder sample (ie. larger FM grains and smaller AF grains) for two key reasons:  1) earlier work by Sarkar et al. has already demonstrated that in similar concentrations of strain-free LCMO nanopowder ferromagnetic order persists in grain sizes reduced to 15 nm with no resulting phase separation 2) the response of the OS$_{\perp}$ strain field at T$_N$ suggests that the onset of AF order is felt by the lattice in the bulk of the sample and not only a minority fraction of smaller grains. 

Finally, the oxygen stoichiometry near the x$=3/8$ concentration in LCMO is known to be very stable once synthesized.  Even while annealing under high pressure oxygen atmosphere at high temperatures, oxygen content can only be slightly tuned \cite{dabrowki}.  That being said, we cannot entirely rule out effects from defect chemistry playing a role in the phase behavior of our LCMO nanopowder\cite{malavasi}; however to the best of our knowledge no report has shown that oxygen nonstoichiometry can stabilize A-type AF order deep within the FM phase in the LCMO phase diagram.  Furthermore, the disappearance of the magnetic phase separation after a rapid hot-press heating process in our annealed sample renders the influence of macroscopic chemical phase separation highly unlikely.  Instead, the removal of the anisotropic strain during the annealing process demonstrates that strain seems to play the dominant role in the stabilization of the phase separated AF order.

\section{Conclusions}
Utilizing strain effects to stabilize coexisting AF order in nanocrystalline LCMO samples presents a novel route for the exploration of phase separation's role within CMR where strain induced correlated magnetic behavior can be directly probed via neutron scattering techniques.  The preferential straining of the b-axis and resulting changes in Mn-O-Mn bond lengths in nanocrystalline samples suggest that phase separation in LCMO can be systematically explored through directly tuning the out-of-plane AF superexchange interactions in competition with the ferromagnetic double exchange. The work presented here motivates future experiments comprehensively exploring this effect in further CMR nanocrystalline materials.

\acknowledgments{
We thank R. Johnson and S. Disseler for help with magnetization measurements. The work is funded by the U.S. National Science Foundation DMR-1056625 (SDW) and the US Department of Energy under contract number DOE DE-FG02-00ER45805 (ZFR). Part of this work was performed at ORNL's HFIR, sponsored by the Scientific User Facilities Division, Office of Basic Energy Sciences, U.S. Department of Energy.}


\end{document}